\input{aipcheck}

\documentclass[
    ,final            % use final for the camera ready runs
  ]
  {aipproc}

\layoutstyle{6x9}

\def\apj{ApJ}
\def\aap{A\&A}
\def\mnras{MNRAS}
\def\araa{ARA\&A}
\def\pasp{Pub. Astr. Soc. Pac.}
\def\rmp{Rev. Mod. Phys.}

\def\mjup{M_{Jup}}
\def\msol{M_\odot}
\def\simgr{\,\hbox{\hbox{$ > $}\kern -0.8em \lower 1.0ex\hbox{$\sim$}}\,}
\def\simle{\,\hbox{\hbox{$ < $}\kern -0.8em \lower 1.0ex\hbox{$\sim$}}\,}

\begin{document}

\title{The mass-radius relationship from solar-type stars to terrestrial planets: a review}

%<Replace this text with PACS numbers; choose from this list: \texttt{http://www.aip..org/pacs/index.html}>}

%\author{<author1>}{
 % address={<author1 address>}
%*}

\author{G. Chabrier}{
  address={ENS-Lyon, CRAL, 69364 Lyon Cedex 07, France}
}

\author{I. Baraffe}{
  address={ENS-Lyon, CRAL, 69364 Lyon Cedex 07, France}
}

\author{J. Leconte}{
  address={ENS-Lyon, CRAL, 69364 Lyon Cedex 07, France}
}

\author{J. Gallardo}{
  address={Departamento de Astronomia, Universidad de Chile, Santiago, 
Chile}
}

\author{T. Barman}{
  address={Lowell observatory, Planetary Research Center, Flagstaff, USA}
%  ,altaddress={<author1 address>} % additional visiting address
}

\begin{abstract}
In this review, we summarize our present knowledge of the behaviour of the mass-radius relationship from solar-type stars down to terrestrial planets, across the regime of substellar objects, brown dwarfs and giant planets. Particular attention is paid to the identification of the main physical properties or
mechanisms responsible for this behaviour. Indeed, understanding the mechanical structure of an
object provides valuable information about its internal structure, composition and heat content as
well as its formation history. Although the general description of these properties is reasonably well
mastered, disagreement between theory and observation in certain cases points to some missing
physics in our present modelling of at least some of these objects. The mass-radius relationship in the overlaping
domain between giant planets and low-mass brown dwarfs is shown to represent a powerful diagnostic
to distinguish between these two different populations and shows once again that the present IAU distinction between
these two populations at a given mass has no valid foundation.
\end{abstract}

\maketitle

%%%%%%%%%%%%%%%%%%%%%%%%%%%%%%%%%%%%%%%%%%%%
%% MAINMATTER
%%%%%%%%%%%%%%%%%%%%%%%%%%%%%%%%%%%%%%%%%%%%

\section{Introduction}

The mass-radius relationship (MRR) of a body in hydrostatic equilibrium at a given time of its
evolution is entirely determined by (i) the thermodynamic properties of its internal constituents, (ii) its
ability to transport and evacuate its internal entropy content, a consequence of the
first and second principles of thermodynamics. In this review, we examine our present understanding
of such properties, in the stellar, substellar and planetary domains, respectively, by comparing
state-of-the-art theoretical calculations with observational determinations of the MRR.

\section{General behaviour of the Mass-radius relation}

Figure 1 portrays the general behaviour of the MRR from the Sun down to gaseous
planets, i.e. over 3 orders of magnitude in mass. The essential physics characteristic of such a behaviour can be
grasped with the help of the polytropic mass-radius relation, $R\propto M^{{1-n\over 3-n}}$ \cite{BurrowsLiebert93,CB00}. The low-mass star (LMS)
regime, from the Sun to the hydrogen-burning minimum mass (HBMM) is characterized by an
evolution of the polytropic index from a value $n\approx 3$, characteristic of
stars with a large radiative core, to $n=3/2$ below $M\sim 0.4\msol$, when the star
becomes fully convective. In the stellar regime, both the ions and the electrons obey classical statistical physics,
so that the combination of a classical nearly perfect gas equation of state (EOS) and of the
quasistatic equilibrium condition yields a
$R\propto M$ dependence. As the mass decreases, the density increases and the internal temperature
decreases, so that the electrons become degenerate (degeneracy can be characterized by a
dimensionless number $\theta= T/T_F$, where $T_F\simeq 3.0\times 10^5 (\rho/\mu_e)^{2/3}$ K is the electron Fermi temperature, with $\mu_e$ the electron mean molecular weight). The onset
of degeneracy ($\theta < 1$) corresponds to the bottom of the main sequence and to the brown dwarf (BD) domain.
Indeed, BDs
are defined as objects not hot enough to sustain hydrogen fusion in their core and thus are
supported primarily by electron degeneracy. Densities in the interior of BDs, however, are not
large enough for the electrons to be fully degenerate, like in the interior of white dwarfs, where $\theta
\simle 10^{-2}$. Electrons in BD interiors
are only {\it partially} degenerate, with $\theta \sim 10^{-1}$-$10^{-2}$. As density decreases with mass in the substellar regime, the electrostatic contribution from the (classical) ions becomes comparable to the (quantum)
electrons one. This yields the flattening in the MRR illustrated in Fig.1. This growing ion
contribution compared with the electronic one leads to
a decreasing polytropic index, from a value $n\sim 3/2$
near the HBMM, where density is highest, to $n\approx 1$ for Jupiter-mass objects. It is easily seen from the above
polytropic relationship that for $n=1$ the radius does not depend on the mass. Eventually, as the body mass decreases, one reaches the regime of terrestrial planets characterized by homogeneous (constant density)
interiors, i.e. incompressible matter, $n=0$. The polytropic index $n$ is directly related to the EOS, $P\propto \rho^\gamma$, with $\gamma =1+1/n$, and thus to the compressibility
$\chi=(\rho {\partial P\over \partial \rho})_T^{-1}=(\gamma\rho^\gamma)^{-1}$. The decreasing
polytropic index when going from nearly perfect gas dominated stellar objects to terrestrial bodies
thus corresponds in first approximation (neglecting the density variation with mass) to less and less compressible interiors.
We will now examine this general MRR behaviour in the various mass ranges and discuss the
agreement between theory and observation.

%%%%%%%%%%%%%%%%%%%%%%%%%%%%%%%%%%%%%%%%%%%%
%% Sample figure:
%%
%% The option [height=...] scales the picture to the given height,
%% without it it would be printed at its nominal size
%%%%%%%%%%%%%%%%%%%%%%%%%%%%%%%%%%%%%%%%%%%%

\begin{figure}
  \includegraphics[height=.5\textheight]{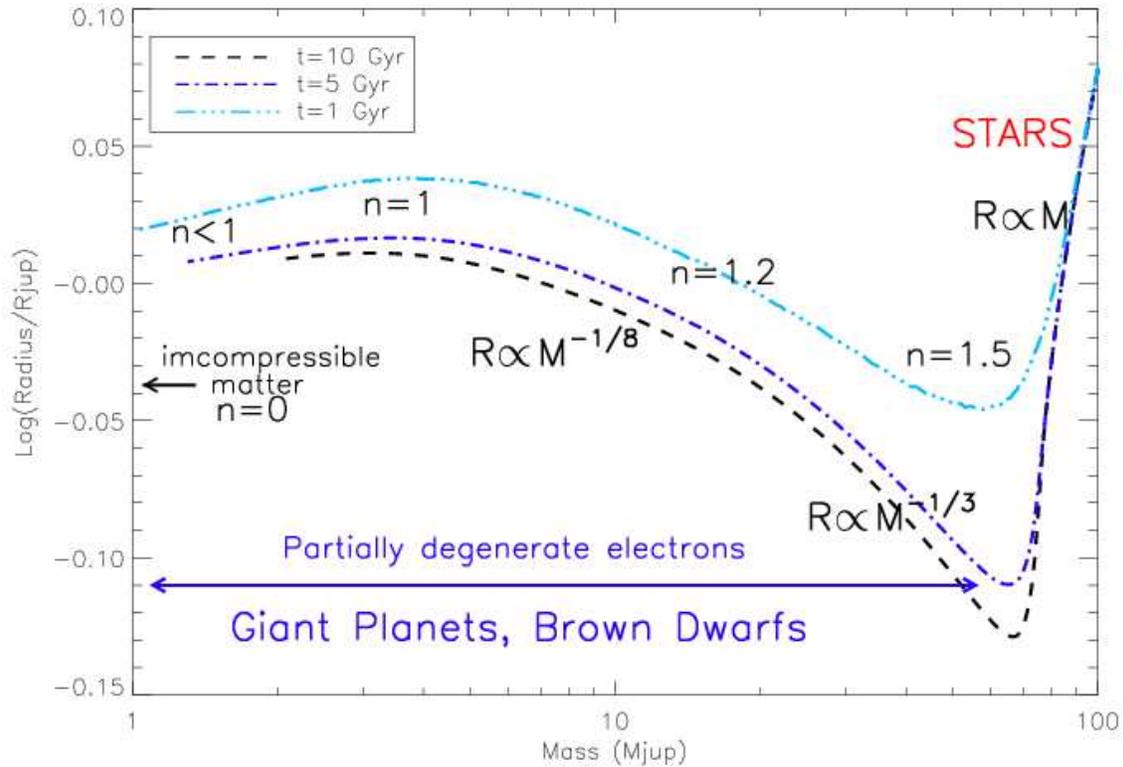}
  \caption{Mass-radius relationship from the Sun to Jupiter for three different isochrones. Characteristic values of the polytropic index $n$ are indicated.}
\end{figure}

\section{The low-mass star range}

The radii of many LMS have been accurately determined from various techniques. Eclipsing binaries
provide the most natural method but include only
a limited number of systems below 1 $\msol$ (see \cite{Lopez07} for a summary of present observational
determinations). Interferometry allows a precise determination of the radii of nearby binaries \cite{Seg03, Berger06},
while transit observations from the OGLE microlensing survey improve significantly the statistics. Comparison between the theoretical and observed
radii from the Sun to the hydrogen-burning limit \cite{Pontetal05} is illustrated in Fig. 1 of \cite{CGB07}. As shown on this figure, the excellent agreement
between theory and observation all along the LMS domain for all the stars except the eclipsing binaries gives confidence in the underlying physics used to determine the mechanical structure of these
cool and dense objects. 
%This agreement has been confirmed recently by ??.
Recent determinations \cite{Casa08} have cast doubt on such a general agreement between theory and observation.
Note, however, that these results are based on {\it indirect} determinations, based on emprical $T_{eff}$-color scales whose accuracy is questioned by some of
the authors
themselves (Bessell, private communication), and cannot be considered as robust as the aforementioned
direct radius  determinations.
The observed radii for the eclipsing binaries, on the other hand, are found to be
systematically $\sim 10$\% larger than all the other {\it observational} determinations, and thus
disagree with the theoretical values at the same level \cite{TorresRibas02}. Problems with the atmospheric opacity have sometimes
been 
invoked as the source of the discrepancy. As shown in Table 1 of \cite{CB97}, however, opacity has
a modest impact on the stellar radius for these compact stars. Changing the metallicity in the atmosphere by a factor
10 (100) affects the radius by a factor $\sim 3\%$ ($\sim 7\%$), so that the opacity of eclipsing binaries should have to be increased to an unrealistic level
to yield the observed 10\% effect on the radius. Missing opacity thus seems to be unlikely to explain the radius discrepancy.
All these eclipsing binaries, however, are fast rotators and are
magnetically very active. It is thus natural to imagine that spot area covers a significant fraction of their irradiating surface, possibly with a modest but non-zero temperature constrast. Chabrier et al. (2007) \cite{CGB07} have suggested that inhibition of internal convection, due to
rotation and/or magnetic field, and/or spot coverage yields a reduction of the internal heat flux and thus a smaller contraction during evolution, providing an appealing explanation for the larger radius in rapidly rotating,
very active stars. The value of the equilibrium field inferred in these (phenomenological) calculations to hamper convection is consistent with the observationally
determined value \cite{ReinersBasri08} and with the one obtained with 3D resistive MHD simulations \cite{Browning08}. As mentioned above, surface density increases with decreasing mass in the LMS regime, reaching a maximum near the
HBMM  \cite{CB00}, so that convection becomes more and more efficient with decreasing mass in this regime.
The aforementioned decreasing convective efficiency is thus expected to be relatively less and less consequential as one moves
along the mass sequence from the Sun to the bottom of the
main sequence. Such a behaviour is indeed supported by observations. These very same effects of magnetically driven inhibition of convection and spot
coverage are also shown to provide a plausible explanation for the temperature revearsal observed in
eclipsing brown dwarfs \cite{Stassun06}, with the most massive object being more affected by magnetic
fields than the smaller one \cite{CGB07}. Interestingly, $H_\alpha$ emission has been detected in the primary of this system at a
 7 times stronger level than the emission from the secondary \cite{Reiners07}.  This brings support to the aforementioned scenario.

Activity in low-mass stars is presently a thriving domain of research. The remarkable results
recently obtained with spectro-polarimetry \cite{Donati08, Morin08} have brought evidence for an evolution of the
topology of the magnetic field with decreasing mass, ie decreasing effective temperature. Whereas objects above about 0.4$\msol$ exhibit a dominantly
toroidal field, objects  below this mass are dominated by a
poloidal, mainly axisymmetric field, although exhibiting only a very modest level of differential rotation.
Although several theoretical calculations have recently suggested various mechanisms to generate
large-scale magnetic fields in fully convective objects like LMS, based on either mean field theories \cite{ChabrierKueker06} or MHD
simulations \cite{Dobbler06}, none of these theories so far can explain the observational results, in particular the
presence of a strong dipolar field in an object with very low level of differential rotation. The recent 3D
resistive MHD simulations of Browning (2008) \cite{Browning08}, although still retaining some limitations, seem to offer a promising avenue to explore this complex
problem. In this
review, we speculate that the abrupt change of topology of the field with decreasing mass,
and the strong decrease of angular momentum loss rate in the same mass range  \cite{ReinersBasri08,Scholz}, are both connected
with the evolution of the Rossby number in  LMS interiors \cite{MohantyBasri03,ChabrierKueker06}. Indeed, the convective
time strongly increases with decreasing mass (i.e. temperature), so that the Rossby number eventually reaches a critical value $Ro\simle 0.1$,
affecting the magnetic field generation process \cite{Donati08, Morin08}. Work to explore this issue is presently under progress.
%Some url test \url{http://www.world.universe}.

\section{The planetary regime}

Extrasolar planets are discovered by radial velocity techniques at an amazing pace. The wealth of discoveries now extends from gaseous giants of several Jupiter masses
 to objects
of a few Earth masses. Detailed
models of planet structure and evolution have been computed by different groups \cite{Fortney07, Baraffe08, Burrows07}. These calculations include various internal compositions, based on presently available high-pressure equations of state for materials typical of planetary interiors. While the models of \cite{Fortney07, Burrows07}
only consider the impact of heavy material on the hydrostatic structure of the planet and neglect the {\it thermal} contribution of these elements to
the planet's cooling rate, the models of \cite{Baraffe08} fully consistently account for this contribution. A detailed discussion
and a comparison of these
models can be found in Baraffe et al. (2008) \cite{Baraffe08} \footnote{Models are availble at http://perso.ens-lyon.fr/isabelle.baraffe/PLANET08/}. This paper also explores the effect of the location
of the heavy element material in the planet, either all gathered at depth as a central core or distributed throughout the gaseous H/He envelope, on the
MRR. It is shown that these different
possible distributions of heavy element can bear important effects on the planet radius and evolution.
Unfortunately, although the average internal composition of the planet
can be inferred with these models from the observed mass and radius, present uncertainties in the EOS of the
aforementioned heavy elements in the T-P range characteristic of planetary interiors prevent a
detailed determination of the internal composition in terms of various heavy element mass fractions. This paper also shows
that the presence of even a modest gaseous (H/He) atmosphere hampers as well an accurate determination
of the internal composition, as the highly compressible gas contains most of the entropy of the planet and thus governs its
cooling and contraction rate. 

This is no longer true for so-called Super-Earth or Earth-like planets, ie objects below the expected limit
for the planet to retain a gaseous atmosphere by gravitational instability, about 10 Earth-masses \cite{Mizuno80, Stevenson82, Rafikov06}. The structure of these "terrestrial planets" has been examined by various groups
\cite{Valencia07, Seager07, Sotin07}. For these objects, although uncertainties in the EOS still prevent precise
determinations of the internal composition, the lack of a substantial gaseous atmosphere allows a more
detailed exploration of the internal composition, opening up the route to accurate determinations of the
composition of exo-Earths as high-pressure experiments of the relevant materials become available.

For about 40 of these systems, the planet is transiting its host star,
allowing a determination of its radius and thus, in combination with the radial velocity observations, of
its mean density. This in turn yields a strong constraint on its average internal composition. Although
planetary evolution models taking into account the effect of the incoming stellar flux on the internal
planetary heat content successfully reproduce the observed radii in many cases \cite{Chabrieretal04, Baraffe05, Burrows07, Baraffe08}, a substantial fraction of these transiting planets exhibit
radii larger than the theoretical determination by a significant amount. Several explanations have
been suggested to explain this puzzling result. Bodenheimer et al.\cite{Boden} invoked tidal heating due to an
undetected companion, a now excluded possibility for HD 209458 b \cite{Laughlin05}, the most illustrative of these abnormalously large planets.
Ongoing tidal heating due to these planets being trapped in Cassini states with large obliquity \cite{WinnHolman05} have been shown to be highly improbable for
short-period planets \cite{Levrard07}.
 Showman \& Guillot \cite{ShowmanGuillot02} suggested
the outer kinetic energy due to the strong stellar irradiation being transported downward and transformed at depth into thermal energy, leading to a hotter isentrope.
The identification of a robust mechanism for
transporting this energy deep enough, however, is still lacking and an accurate (so far missing) description of the (probably small-scale) dissipative processes in such
natural heat engines is mandatory to assess the validity and the importance of this mechanism for hot-Jupiters
\cite{Goodman08}. Burrows et al.\cite{Burrows07} arbitrarily invoke a combined effect of tidal heating, in some cases, and/or strongly
enhanced (10 times) atmospheric opacity as a possible explanation. These calculations, however, do
not provide any explanation for such a persisting strong metal enrichment in the planet's {\it radiatively stable}
atmosphere and outer envelope, where gravitational sedimentation should occur. Moreover, these calculations do
not consider any increase of molecular weight due to such a heavy element enrichment in the envelope. Similar calculations by Guillot \cite{Guillot08}
including consistently the increase of mean molecular weight in the envelope show that, in most cases, such an increase of heavy element abundance in the planet's atmosphere leads to a {\it smaller} radius. More recently, Chabrier \& Baraffe (2007) \cite{CB07} suggested that the onset of
layered or oscillatory convection, due to the presence of an internal compositional gradient, may hamper
the internal heat flux transport, slowing down the planet's contraction. Although such layered convection is observed in many situations
in Earth lakes or oceans, due to the presence of salt concentrations (the so-called thermohaline convection), it remains unclear whether this process can occur under giant
planet interior conditions. Interestingly, although
the Showman \& Guillot scenario necessarily needs the planet to be strongly irradiated, the Chabrier \& Baraffe one does
not, even though irradiation does favor the onset of layered convection. The observation of an inflated
transiting planet far enough from its parent star for the incident flux to have a negligible effect on the planet's
internal heat content would provide a clear demonstration of the validity of the Chabrier \& Baraffe scenario. Kepler or Corot will hopefully
provide such observational diagnostics.

\section{The overlaping domain: from planets to brown dwarfs}

The distinction between BDs and giant planets has become these days a topic of intense debate. In 2003,
the IAU has adopted the deuterium-burning minimum mass, $\sim 10\,\mjup$,
 as the official distinction between the two types of objects.
We have discussed this limit in previous reviews \cite{Chabrier03, Chabrieretal05, Chabrieretal07} and shown
that it does not rely on any robust physical ground and is a pure semantic definition.
The observation of free floating objects with masses of the order of a few jupiter masses in (low extinction) young clusters
\cite{Caballero07} shows that star and BD formation extends down to Jupiter-like masses,
with a limit set up most likely by the opacity-limited fragmentation, around a few Jupiter-masses \cite{BoydWhitworth05}.
%Since brown dwarfs and planets are likely to form through different mechanisms. 
Observations show
that young brown dwarfs and stars share the same properties and are consistent with BDs and stars
sharing the same formation mechanism \cite{Andersen08, Joergens08} (for a recent review see \cite{Luhmanetal07}). On the other hand, the fundamentally different mass distribution of exoplanets detected by radial velocity surveys \cite{Udry08} clearly suggests a different formation
mechanism, consistent with the so-called core accretion scenario \cite{Pollack96, Alibert05}. Consequently, planets are believed to have a substantial enrichment in heavy elements compared with their parent star, as observed for our own solar giant planets, whereas BDs of the same
mass should have the same composition as their parent cloud, ie a  $Z\sim 2\%$ heavy element mass
fraction for a solar environment.

\begin{figure}
  \includegraphics[height=.5\textheight]{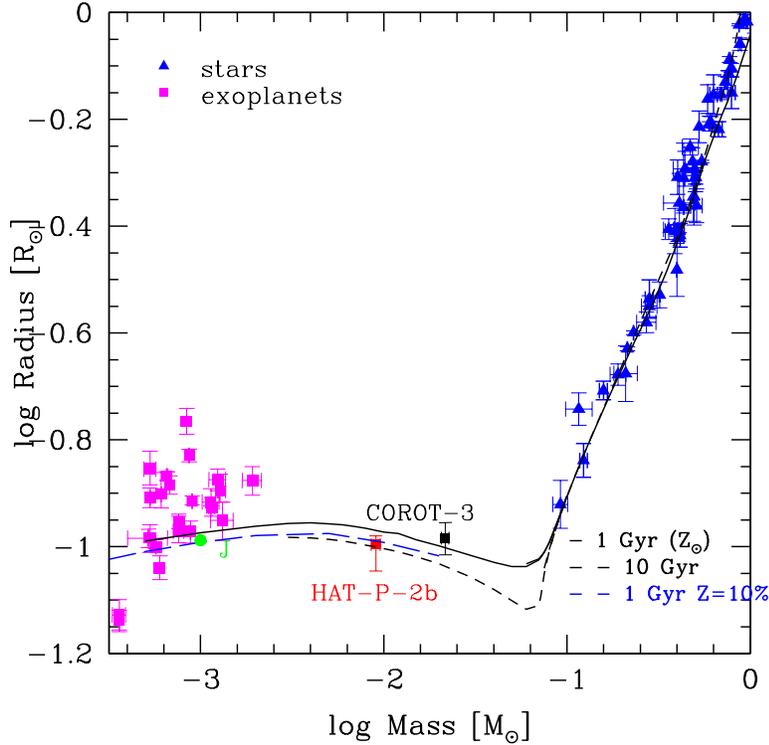}
  \caption{Mass-radius relationship from from the stellar to the planetary regime. The (black) solid and short-dash lines correspond to models with solar composition,
for two isochrones. The (blue) long-dahs line corresponds to an object with a $Z=10\%$ mass fraction of heavy elements (from \cite{Baraffe08}).
The observationally-determined values of Hat-P-2b and Corot-3b are indicated.}
\end{figure}

As shown in the previous section, an internal heavy material enrichment yields a smaller radius, for a given mass, and
thus the MRR provides in principle a powerful diagnostic to distinguish planets
from BDs in their overlaping mass domain. Figure 2 portrays the MRR in the substellar
regime, with the observationally determined radii of Hat-P-2b \cite{Winn07} and Corot-3b \cite{Deleuil08}. These objects
provide the first observational constraint in the mass-range between Jupiter-mass planets and the HBMM, where the MRR is predicted to follow the behaviour illustrated in Fig. 1.
The first important result is that the observations do confirm the theroretical predictions, providing
confidence in the description of the internal physics characteristic of the cool, dense, partially degenerate objects known as "substellar objects". A detailed composition of Hat-P-2b has been derived
in Baraffe et al.(2008) \cite{Baraffe08}. Models are shown in Fig. 2 for two different internal compositions: one, with a solar abundance of heavy
elements, corresponds to solar-metallicity BDs,for two isochrones, whereas the other one, with a 5-times solar metal enrichment, corresponds to massive gaseous planets. Assuming that (i) the theoretical
MRR is accurate, (ii) the observational error bars on the radius are reliable, the second
important result illustrated in Fig. 2 is that, given the age inferred for the system, $\sim 2$-3 Gyr \cite{Bakos07}, Hat-P-2b is too dense to be a BD. This in turns shows that
planets can form up to at least 9 Jupiter masses, a result of prime importance for constraining models of planet formation. Although such a mass is still compatible
with planet formation models
based on the core-accretion scenario \cite{Mordasini07}, an alternative possibility for the formation of such high-mass,
short-period planets is collisions between less massive planets (see \S 6 of \cite{Baraffe08}).
Interestingly, although Corot-3b is compatible with this object being a BD with solar composition, for the correct age of the system, $\sim 2$ Gyr \cite{Deleuil08},
 one cannot exclude this object to be a strongly inflated irradiated planet with a massive core. Work is under progress to
examine this possibility in more details.

\section{Conclusion and perspectives}

As shown in this review, the non-monotonic behaviour of the mass-radius relationship from the
stellar to the planetary regime, through the brown dwarf domain, can be understood qualitatively
and quantatively in terms of the physical properties of the ionic and electronic fluids under the
appropriate conditions. It is interesting to see that these theoretical predictions have been confirmed by the
subsequent observational determinations from the Sun down to the domain of giant planets.
The evolution of active objects, both in the stellar and substellar regime, is suggested to differ from
the one of non-active objects, as both magnetic field and fast rotation are predicted to affect the
internal heat transport and/or escaping flux. A quantitative assessment of this point, however, is
still lacking and requires 3D resistive MHD numerical simulations over pressure scale heights characteristic of
fully or dominantly convective objects, a formidable challenge. In the same vein, observations have
shown that the topology of the magnetic field in LMS interiors varies abruptedly around about 0.4 $\msol$,
near the expected transition from centrally radiative to fully convective stars. It is not clear yet
what is the main reason for such a strong variation but the Rossby number seems to play a key role in this process.

In the planetary domain, evolution models incorporating EOS for various materials appropriate for
planetary interiors, and taking consistently into account the thermodynamic contribution of such materials
both on the structure and on the cooling of the planet have become available and provide a reliable 
diagnostic to infer the internal composition of these planets. Although the presence of a gaseous
atmosphere only allows the determination of the planet's gross internal composition, a more detailed
balance between the various components can be obtained for terrestrial planets, although solutions remain degenerate.
At any rate, all these internal composition determinations provide
strong constraints on the formation mechanism for gaseous, icy and terrestrial planets. All these
determinations are consistent with the core-accretion model for planet formation. Conversely, the
large heavy material enrichment inferred for many of these planets clearly excludes the
gravitational instability scenario. The only remaining, although uncertain possibility for this latter is the formation of 
planets at very large distances ($\simgr 100$ AU), for the disk, assuming it is massive enough, to be cold enough
to violate the Toomre stability condition \cite{Rafikov05,Whitworth06} (see \cite{Dullemond08} for a recent review).
The puzzling inflated radius observed for many transiting planets
still remains unexplained, and very likely points to some missing physical mechanism in the description of these objects.
The expected wealth of transiting planets at large ($\simgr 0.1$ AU) orbital distances from COROT and KEPLER will
hopefully enable us to solve this intriging problem. 

The recent observation, by radial velocity and by the COROT mission, of transiting objects around 10 $\mjup$,
in the overlaping mass range between planets and brown dwarfs, confirms the theoretical m-R
relationship and opens the door to an observational diagnostic to distinguish brown dwarfs and planets
and thus to determine, in a foreseable future, the minimum mass for star formation and the maximum mass for planet formation. The theoretical exploration of these observations already suggests that
planets should form up to masses of at least 9 $\mjup$.

%%%%%%%%%%%%%%%%%%%%%%%%%%%%%%%%%%%%%%%%%%%%%%%%
%% BACKMATTER
%%%%%%%%%%%%%%%%%%%%%%%%%%%%%%%%%%%%%%%%%%%%%%%%

\begin{theacknowledgments}
This work was supported by the french "Agence Nationale pour la Recherche (ANR)" within
the 'magnetic protostars and planets (MAPP)' project and by the "Constellation" european network MRTN-CT-2006-035890.
\end{theacknowledgments}

%%%%%%%%%%%%%%%%%%%%%%%%%%%%%%%%%%%%%%%%%%%%%%%%
%% The bibliography can be prepared using the BibTeX program or
%% manually.
%%
%% The code below assumes that BibTeX is used.  If the bibliography is
%% produced without BibTeX comment out the following lines and see the
%% aipguide.pdf for further information.
%%
%% For your convenience a manually coded example is appended
%% after the \end{document}
%%%%%%%%%%%%%%%%%%%%%%%%%%%%%%%%%%%%%%%%%%%%%%%%

%%%%%%%%%%%%%%%%%%%%%%%%%%%%%%%%%%%%%%%%%%%%%%%%
%% You may have to change the BibTeX style below, depending on your
%% setup or preferences.
%%
%%
%% For The AIP proceedings layouts use either
%%%%%%%%%%%%%%%%%%%%%%%%%%%%%%%%%%%%%%%%%%%%

\bibliographystyle{aipproc}   % if natbib is available
%\bibliographystyle{aipprocl} % if natbib is missing

%%%%%%%%%%%%%%%%%%%%%%%%%%%%%%%%%%%%%%%%%%%
%% You probably want to use your own bibtex database here
%%%%%%%%%%%%%%%%%%%%%%%%%%%%%%%%%%%%%%%%%%%
%\bibliography{sample}

%%%%%%%%%%%%%%%%%%%%%%%%%%%%%%%%%%%%%%%%%%%
%% Just a reminder that you may have to run bibtex
%% All of it up to \end{document} can be removed
%% if you don't like the warning.
%%%%%%%%%%%%%%%%%%%%%%%%%%%%%%%%%%%%%%%%%%%
%\IfFileExists{\jobname.%bbl}{}
% {\typeout{}
%  \typeout{******************************************}
%  \typeout{** Please run "bibtex \jobname" to optain}
%  \typeout{** the bibliography and then re-run LaTeX}
%  \typeout{** twice to fix the references!}
%  \typeout{******************************************}
%  \typeout{}
% }

%%%%%%%%%%%%%%%%%%%%%%%%%%%%%%%%%%%%%%%%%%%
%% The following lines show an example how to produce a bibliography
%% without the help of the BibTeX program. This could be used instead
%% of the above.
%%%%%%%%%%%%%%%%%%%%%%%%%%%%%%%%%%%%%%%%%%%

\end{document}